\documentstyle[12pt]{article}
\def \be {\begin{equation}}
\def \eq {\end{equation}}
\def \bee {\begin{eqnarray}}
\def \eqq {\end{eqnarray}}
\def \n {\nonumber}
\def \bea {\begin{array}{c}}
\def \eqa {\end{array}}

\def \CN {{\cal N}}

\def \CL {{\cal L}}
\def \CE {{\cal E}}
\def \BR {{\bf R}}
\def \Z {{\bf Z}}

\def \Ds {{D \hspace{-7.4pt} \slash}\;}
\def \ds {{\del \hspace{-7.4pt} \slash}\;}
\def \da {\dagger}
\def \gy {g_{YM}}
\def \a {\alpha}
\def \b {\beta}
\def \g {\gamma}

\def \s {\sigma}
\def \del {\partial}
\def \th {\theta}

\def \ad  {{\dot{\alpha}}}
\def \bd  {{\dot{\beta}}}
\def \ev {{\bf{8_v}}}
\def \es {{\bf{8_s}}}
\def \ec {{\bf{8_c}}}

\begin{document}
\begin{titlepage}
\begin{center}
\hfill    SISSA 68/97/EP/FM\\
\hfill    hep-th/9705137\\
\vskip .2in
{\LARGE {\bf On the String Interpretation of M(atrix) Theory}}
\vskip .2in
{ \large L. Bonora$^{a,b}$, C.S. Chu $^{a}$}
{}~\\
\quad \\
{\em ~$~^{(a)}$ International School for Advanced Studies (SISSA/ISAS),}\\
{\em Via Beirut 2, 34014 Trieste, Italy}\\
{\em ~$~^{(b)}$  INFN, Sezione di Trieste}\\
\vskip .1in
E-mail: bonora@sissa.it, cschu@sissa.it
\end{center}
\begin{abstract}
It has been proposed recently that, in the framework of M(atrix) theory,
$\CN=8$ supersymmetric $U(N)$ Yang-Mills theory in 1+1 dimensions 
gives rise to type IIA long string configurations.
We point out that the quantum moduli space of $\mbox{SYM}_{1+1}$ gives 
rise to two quantum numbers, which fit very well into the M(atrix) theory.
The two quantum numbers become familiar if one switches to a IIB picture,
where they represent configurations of D--strings and fundamental strings.
We argue that, due to the $SL(2,\Z)$ symmetry, of the IIB theory, such
quantum numbers must represent configurations that are present also in the 
IIA framework.
\end{abstract}
\end{titlepage}

\section{Introduction}

It has recently been proposed \cite{motl,BS,DVV1}, in the framework of
M(atrix) theory \cite{BFSS}, that IIA string theory in the light cone
can be identified with the two dimensional $\CN=8$
supersymmetric $U(N)$ Yang-Mills theory ($\mbox{SYM}_{1+1}$) 
in the $N\to\infty$ limit. This IIA theory contains not only the ordinary
string configurations, but also new non--trivial string 
configurations, the `long strings' \cite{DMVV}, which naturally arise from the
$\mbox{SYM}$ moduli space. To stress the distinction with the usual type IIA
string theory, at times we refer to the latter theory as 
the `enlarged IIA theory'.

In this paper we point out that the `quantum moduli space' of a  
$\mbox{SYM}_{1+1}$ theory is characterized by a couple of quantum numbers 
$(m,n)$ that fit very nicely in the framework of the compactified M(atrix) 
theory and explain not only the `long strings' but also other topological
configurations that are traditionally not present in a IIA picture. The latter
become more natural if, by a T--duality operation, we switch to a IIB 
picture. In a IIB context the two quantum numbers correspond to configurations
of D--strings and elementary strings and it is, of course, natural for all 
of them to feature in the theory, due to the $SL(2,\Z)$ symmetry of the latter.
Thus, to the extent that M(atrix) theory is an exact description of M theory,
we must supplement IIA theory not only with long strings, but also with 
additional topological configurations (which will be described below).

In this regard we can also add another remark. The `long string' configurations
in \cite{motl,BS,DVV1} were included in the spectrum as they represent
the twisted sectors of the $\mbox{SYM}_{1+1}$ orbifold moduli space, 
as is usual in ordinary string orbifold constructions in order
to guarantee modular invariance. The results of this paper, 
briefly stated above, show that the role analogous to `modular invariance'
for the enlarged IIA theory is played by the $SL(2,\Z)$ symmetry of the 
IIB theory. 

The paper is organized as follows. In the next section we use well--known
results on Yang--Mills theories in 1+1 dimensions, to give an explicit
representation of the `quantum moduli space' for these theories, as 
anticipated above. In 
section  3, we apply this notion to the M(atrix) framework. Section 4 is 
devoted to some comments.

\section{SYM Theory on a Cylinder}

In this section we use well--known results on Yang--Mills 
theory in 1+1 dimensions,
and reformulate its moduli space in a way which is suitable 
to the present context.
Let us consider a theory,
with gauge group $G=U(N)$, defined on a circle of circumference 
$L= 2 \pi R$ and suitably 
rescaled so that the pure Yang--Mills part becomes
\be \label{YM}
S =  \frac{1}{4\gy^2}\int ~dtd\s~ tr F_{\mu \nu}^2  ,
\eq
where $\gy$ is the gauge coupling constant. 
  
The theory (\ref{SYM}) admit the boundary conditions
\be\label{bound}
\phi(t,\s+L) = U \phi(t,\s)U^\dagger,
\eq
where $\phi$ is any of the matrix--fields involved and $U$ is a constant
matrix in $U(N)$. 
We first notice that  if $U$ is in the center of 
the gauge group,  then the boundary condition (\ref{bound}) corresponds to 
periodicity. Otherwise, a generic $U$ 
corresponds to a theory over a noncompact space, unless 
$U^k\in U(1)$ -- which without loss of generality can be reduced to
$U^k={\bf 1}$ -- for some minimal $k$. We are not interested in the generic 
case, while our attention will concentrate on the latter 
case which corresponds
to a gauge theory defined (i.e. with periodic boundary conditions) on a
$k$--th covering of the circle, that is a circle of radius $kR$. We
conventionally refer to this circle as a `large circle' when $k>1$,
as opposed to the original `small circle' of radius $R$. This splits the 
theory into different sectors, which, from the point of view of 
the gauge fixing, can be treated separately. In each sector 
the allowed gauge transformations are the periodic ones on the corresponding
large circle. 

Any $U$ satisfying $U^k={\bf 1}$ for some finite $k$ belongs to some
finite discrete subgroups of $U(N)$, but, in the following, we restrict 
ourselves to those $U$ that belong to the Weyl group $S_N$ (the permutation 
group of $N$ objects) of $U(N)$.
\footnote {It is an open question whether
$U$'s not belonging to $S_N$ may generate interesting 
physical configurations like the `long strings' of the following section.}
And to avoid any confusion we
denote the elements of $S_N$ by $g$, instead of $U$.
Each distinct sector is specified, in this case, 
by a conjugacy class of $S_N$, denoted as $[g]$. We recall that a 
conjugacy class $[g]$ of $S_N$ corresponds to a partition of $N$ into `cycles'. 
It is a theorem of finite group theory that the order $k$ of $[g]$ is 
the least common multiple of the degrees of the cycles contained in 
the partition.

   We recall that, in each sector, we have periodic boundary conditions over the
appropriate large circle, and the gauge fixing can therefore be
carried out as in \cite{hos,sem,hetrick}. 
One first solves the holonomy equation $(\del_1 - A_1)V=0$ and, by means 
of the solution $V$, constructs a gauge transformation that 
maps $A_1(t,\s)$ to a $\s$--independent potential; then, by means of another
$\s$--independent gauge transformation one brings $A_1$ to the diagonal form.
This entails also $A_0=0$. In summary
\footnote{ 
In fact, the gauge fixing also implies that, if
${A_0(t,\s)}_{ij} =  \sum_n {a_n(t)}_{ij} e^{2 \pi i n \s /L}$,
then ${a_n(t)}_{ij}$ can be nonvanishing in some particular cases. 
In any case the ${a_n(t)}_{ij}$'s appear in the action as Lagrange multipliers
that impose conditions which are irrelevant in this paper.}
\be \label{gaugefix}
A_0=0, \quad A_1(t,\s) = \frac{ 1}{L}  \mbox{\, diag \,}
(\b_1(t), \b_2(t),\cdots, \b_N(t)).
\eq
In this gauge, the electric field is given by
\be \label{Efield}
E =  \frac{ 1}{L} \mbox{\, diag \,}
(\bd_1(t), \bd_2(t),\cdots, \bd_N(t))
\eq
and, since the Faddeev--Popov determinant for gauge fixing (\ref{gaugefix}) 
is trivial, the action (\ref{YM}) reduces to
\be \label{OQM1}
S= \int dt \CL, \quad \CL= \frac{ 1}{2 \gy^2 L} \sum_{k=1}^{N} \bd_k^2. 
\eq
It is easy to see that 
the gauge fixed configurations have in general a residual gauge symmetry $\Z^N$:
\be\label{G0}
\b_k \rightarrow \b_k + 2 \pi n_k, 
\quad n_k \in \Z,
\eq
This amounts to restricting the
$\b_k$'s to be living on the torus $T^N$. 

In addition to (\ref{G0}), (\ref{gaugefix}) has another
residual gauge symmetry.
In every sector we can go back from the `large circle' to the 
`small circle' where the theory is originally formulated. In so doing we must
switch on again the boundary conditions (\ref{bound}).
It is easy to see what the allowed boundary conditions and residual gauge 
transformations in each sector are. The boundary condition
\be\label{Abound}
A_1(t,\s+ L) = g A_1(t,\s)g^\dagger
\eq
together with $A_1$ being diagonal, implies that, since $g$ is in $S_N$, 
the only operation it performs is to permute the $\beta_k$ eigenvalues 
among themselves. In relation to the given $g$,
the set of $\beta_k$'s will split uniquely into cycles within 
each of which $g$ acts irreducibly while the cycles are not permuted.
Then, as a consequence of (\ref{Abound}) and the $\s$--independence of $A_1$, 
the $\beta_k$'s in each cycle must all be equal. 

Now, in any sector, the residual gauge symmetry is $S_N$, since acting with
any $h\in S_N$ on $A_1$, will preserve (\ref{Abound}) with $g$ replaced by
$hgh^\dagger$, i.e. will preserve the sector.
The above conclusion is tantamount to saying that the moduli space of the
gauge degrees of freedom of the theory (\ref{SYM}), for each sector, 
is the orbifold 
\be \label{OQM2}
\BR^N /G_N
\eq
where $G_N$ is the semidirect product of $S_N$ and $\Z^N$.

What we had said so far accounts for the second of the two integers announced 
in the introduction. The first one comes from quantization.
We recall that the $\beta_k$'s live on $T^N$. Therefore
the canonical momenta $\pi_k$ conjugate to the $\b_k$'s are quantized,
\be
\pi_k =\frac{1}{\gy^2} E_k, \quad \pi_k \in \Z
\eq
and the energy from the pure gauge sector is
\be \label{noncompact}
\CE = \frac{\gy^2 L}{2} \sum_{k=1}^{N} \pi_k^2.
\eq

As an aside, let us recall that 
this is the so--called non--compact quantization
\cite{hetrick}, as opposed to the compact one \cite{raj} in which
one directly quantizes the Wilson loops.  
In the compact quantization the characters $\chi_R$ of the irreducible
representations $R$ of $U(N)$ form a basis of the Hilbert space of gauge 
invariant functions. The corresponding energy is given by 
\be \label{compact}
\CE_R = \frac{\gy^2 L}{2} C_2(R)
\eq
where 
\be\label{Casimir}
C_2(R)= \sum_{i=1}^N f_i^2 + \sum_{i=1}^N(N+1-2i)f_i 
\eq
is the quadratic Casimir for the representation $R$ specified by 
a Young tableau
with $f_i$ boxes in the $i^{th}$ row.
Comparing the energies in the two quantizations we see that they agree up
to the second term in the RHS of (\ref{Casimir}), which is generated by the
curvature of the group manifold. Therefore one can make the identification
\be \label{rel}
f_i \equiv \pi_i.
\eq

Let us return to the non-compact quantization.
In the case of twisted sectors (\ref{Abound}) for which the $\b_k$'s
within the same cycle have to be identified, one must introduce some
obvious modifications due to such constraints. For example, in a subsector 
given by a cycle of length $n$, one gets
\be \label{pisec}
\pi_k =\frac{n}{\gy^2} E_k, 
\quad\quad \pi_1 = \pi_2 =\cdots = \pi_n \in \Z
\eq  
and
\be \label{Esec}
\CE_{cycle} = \frac{\gy^2 L}{2n^2} \sum_{k=1}^{n} \pi_k^2, 
\eq

The analysis of the moduli space, carried out in this section, is strictly
valid for pure Yang--Mills. The addition of matter may modify its structure.
However this will not happens for the $\CN=8$ supersymmetric theories we 
will consider in the following section. 

Finally, with some abuse of language, we define what we mean by `quantum moduli 
space' of a $\mbox{SYM}_{1+1}$, like the ones considered in the next section. 
A point in it is a collection of subsectors, each of 
which is characterized by two integers $(m,n)$, where $n$ denotes the degree or 
length
of a cycle in a given partition of $N$, and $m$ represents the common eigenvalue
of the $\pi_k$ contained in that cycle. As we will see in the
next section, these subsectors admit different stringy interpretations in
different theories.

\section{IIB versus IIA}

According to the original proposal of \cite{BFSS},
M theory in the infinite momentum frame is described by the
$U(N)$ supersymmetric quantum mechanics $\mbox{SYM}_{1+0}$
in the limit $N \rightarrow \infty$
\bee
L &=& \frac{1}{2} tr \left( 
\frac{1}{R_{11}} \dot{X}^{i 2} - R_{11} [X^i, X^j]^2 -\th^T \dot{\th}
-R_{11} \th^T \g_i [X^i,\th] \right) \n \\
\label{qm}
H&=& \frac{ R_{11}}{2} tr 
\left( \Pi_i^2 +[X^i, X^j]^2 + \th^T \g_i [X^i,\th] \right)
\eqq
$R_{11}$ is the radius of the original longitudinal 
circle on which M theory is compactified. 
For finite $N$, the $\mbox{SYM}_{1+0}$ describes the low energy effective
dynamics (at short distances and low velocities) \cite{ulf}
for a system of $N$ D0 branes with string coupling
$g_s = (R_{11}/l_p)^{3/2}$ and is a partial description
of the compactified M theory, i.e. 10 dimensional IIA string theory.
The total longitudinal momentum is given by
$p_{11}=N/R_{11}$. The limit $N \rightarrow \infty$ takes
one to the infinite momentum frame. 
It was argued that all 
relevant velocities vanish in the same limit; this  together
with a conjectured nonrenormalization theorem for the $v^4$
term in the effective action for D0 brane  
justifies the use of (\ref{qm}) as an exact nonperturbative
formulation for M theory in the infinite momentum frame.

Toroidal compactification of (\ref{qm})
for finite $N$ has been carried out in details in \cite{wati} and
a two dimensional SYM model was obtained.  
For our purpose, we repeat this procedure
in the Appendix and get the following  two dimensional $\CN=8$ 
$\mbox{SYM}_{1+1}$ on a cylinder
$0 \leq \s \leq 2 \pi$,
\bee \label{SYM'}
S &=& \frac{1}{2 \pi l_s^2} \int tr \left(
(D_\mu X^i)^2   + \th^T \Ds \th
+ \frac{1}{\gy^2} F_{\mu \nu}^2
-\gy^2  [X^i,X^j]^2
+ \gy \th^T \g_i [X^i,\th]
\right).
\eqq
The fields $X^i, i=1,\cdots,8$ transform in
the vector representation $\ev$ of
$SO(8)$, while the two  spinors
$\th^\a_L, \th^\ad_R, \a,\ad =1,\cdots, 8$
transform in the representations $\es, \ec$ and have opposite chirality.
All the fields are $N\times N$ hermitian matrices.
The coupling $\gy$ depends on the parameters $R_9$ and $R_{11}$ 
and this dependence can be modified by means of
field redefinitions. Later, we will use this fact to
study the different stringy interpretations of (\ref{SYM'}). 

M theory compactified on a circle is supposed to give 10 dimensional
nonperturbative IIA string theory. 
Given the BFSS formulation for M theory, 
one can get a nonperturbative formulation for IIA
string theory by studying the compactification of (\ref{qm}). 
The prescription of sect. 9 of \cite{BFSS} and \cite{wati} 
is to take the large $N$ limit of
the SYM (\ref{SYM'}) resulting from the compactification of 
(\ref{qm}).
To this purpose, the authors of \cite{DVV1} obtained  the following 
supersymmetric $U(N)$ Yang-Mills theory 
($\mbox{SYM}_{1+1}$) as a nonperturbative formulation for
IIA string theory, 
\be \label{SYM}
S=\frac{1}{2\pi} \int tr \left(
(D_\mu X^i)^2 + \th^T \Ds \th + g_s^2 F_{\mu \nu}^2 -\frac{1}{g_s^2}
[X^i,X^j]^2 + \frac{1}{g_s} \th^T \g_i [X^i,\th] 
\right).
\eq
in units $l_s=1$.
In (\ref{SYM}), the radius $R_{11}$ is 
understood and $N$ is considered in the limit $N\to \infty$. 
The identification with type IIA string theory in \cite{DVV1} is made
via the flip $R_9 \leftrightarrow R_{11}$ so that the string 
coupling $g_s$ is related to $R_9$ by
\be \label{dvvgs}
g_s =(R_9/l_p)^{3/2} = R_9/l_s.
\eq

The action (\ref{SYM}) is not the only possible representation 
of (\ref{SYM'}) we want to discuss. 
In particular,  we consider the following identifications  of $\gy$ 
admitting stringy interpretations: 
\bee 
{\cal A}_D &:& \gy^2 = (l_p/R_{11})^3 /l_s^2,  \label{ad}\\ 
{\cal A}_F &:& \gy^2 = (l_p/R_9)^3 /l_s^2,    \label{af} \\
{\cal B}_D &:& \gy^2 = (R_{11}/R_{9})/{l_s^2} ,   \label{bd}\\
{\cal B}_F &:& \gy^2 = (R_{11}/R_{9})^2/{l_s^2}. \label{bf} 
\eqq 
In the following, however, we set $l_s=1$ throughout.

We now explain what we mean by ${\cal A}_D$, 
${\cal A}_F$ etc.. 
First of all, as long as $N$ is finite, we interpret (\ref{SYM'})  as 
representing a 9--dimensional theory. 
According to the current M(atrix) theory interpretation, 
upon taking the limit $N\to \infty$, 
we are supposed to recover 10 dimensional theories. 
We will comment on this point at the end of the paper.  Throughout 
this section we keep $N$ large but finite.

The most obvious interpretation comes from the very 
construction of the D0--brane theory of BFSS: 
(\ref{SYM'}) represents
a theory of D0--branes with string coupling 
$g_s=R_{11}$ compactified on a circle of 
radius ${R_9}$ in the $9^{th}$ direction. 
Let us call it ${\cal A}_D$. 
Flipping the $9^{th}$ and $11^{th}$ dimensions, 
we obtain a new version of the theory, which we refer
to as ${\cal A}_F$. A suitable action for this theory is (\ref{SYM}). 
This theory, in the $N\to \infty$ limit, 
has been interpreted in \cite{DVV1} as a theory of IIA strings,
see also \cite{motl,BS}.  
On the other hand, starting from ${\cal A}_D$ and 
performing a T--duality operation in the 
$9^{th}$ direction we get a theory of type IIB D--strings, \cite{BFSS,wati}. 
For finite $N$,
we recall that (\ref{SYM'}) was obtained by compactifying 10 dimensional
D0 dynamics on a circle of radius $R_9$, so by T--duality,
(\ref{SYM'}) represents a collection of $N$ D--strings, with
`coordinates' $X^i, \th^i$, stretched in the $9^{th}$ direction \cite{HW} 
on a circle of radius $1/R_9$, with the string coupling
\be \label{bdcoup}
g_B = R_{11}/R_{9}.
\eq
We have chosen the YM coupling (\ref{bd}) 
\be 
\gy^2 =  g_B 
\eq
for this case, so that the limit of  strongly coupled SYM
corresponds to the limit of free D--strings.
We refer to this interpretation as ${\cal B}_D$.
If we now make the $9 \leftrightarrow 11$
flip, as above, we invert the string coupling,
\be \label{bfcoup}
g_B = R_{9}/R_{11}.
\eq
i.e. we do an S--duality 
operation, therefore it is natural to think that we end up in this way 
with a type IIB theory in which fundamental strings 
replace D--strings and vice versa. 
The action (\ref{SYM'}) together with 
\be
\gy^{2} =  1 / g_B^2
\eq
now describes $N$ fundamental strings 
with `coordinates' $X^i, \th^i$. 
The  strongly coupled SYM corresponds to weekly
coupled fundamental IIB strings. 
For this reason we call it ${\cal B}_F$. 
Finally it is to be expected that a T--duality operation in the $9^{th}$ 
direction will map ${\cal B}_F$ back to ${\cal A}_F$. In sum we have the
diagram
\be\nonumber
\begin{array}{ccc}
{\cal A}_D & \leftrightarrow & {\cal B}_D\\
{\updownarrow} &{}&{\updownarrow}\\
{\cal A}_F&\leftrightarrow&{\cal B}_F
\end{array}
\eq
where vertical arrows represent the 9--11 flip, while the horizontal ones
represent the T--duality operation. 
A similar diagram has appeared in \cite{DVV1}.
(\ref{SYM'}) with the specifications (\ref{ad}--\ref{bf}) 
provides a suitable description for all these cases. 

The stringy interpretation of the SYM model (\ref{SYM'})
is clearer by going to the strong $\gy$ limit. 
This limit  corresponds to strongly 
coupled gauge dynamics and is governed by an IR fixed point theory.
In this limit, the fields $X$ and  $\th$ become diagonal 
and for this reason, a stringy interpretation is apparent.  
In (\ref{ad}-\ref{bf}), we have arranged the $\gy$ so that
this limit corresponds to an appropriate string picture in each case. 
It was argued in \cite{ST,DVV1} 
that there is a nontrivial identification by the symmetric group
$S_N$ and the IR theory is the $\CN=8$ supersymmetric sigma model
conformal field theory  
\be \label{OCFT1}
S=\frac{1}{2\pi} \int tr \left(
(\del_\mu X^i)^2 + \th^T \ds \th 
\right)
\eq
with the orbifold target space
\be \label{OCFT2}
S^N \BR^8 = (\BR^8)^N/S_N.
\eq 
The two spinors $\th_L$ and $\th_R$
have opposite $SO(8)$ chirality, which is the chirality setting for  
IIA strings in the light cone gauge. This fact was used by
\cite{motl,BS,DVV1} to identify  the untwisted sector 
of the orbifold field theory 
with $N$ IIA strings in the light cone gauge, while 
twisted sectors give IIA strings of different lengths.  

The above interpretations are rather natural; however, as we have already
pointed out, when talking about type
IIA or IIB, we do not refer simply to the old type IIA and IIB string theories,
but to enlarged theories which contain the old string theories as a 
particular subsector. The rich structure of these theories come directly
from the analysis of the $\mbox{SYM}$
model, underlying all of them, which we have
done in the last section. In general all four theories split into subsectors
labeled by two integers $(m,n)$. They come either from the quantization
of the boundary conditions (the cycles of the previous section) or
from the quantization of the gauge potential degrees of freedom 
(the common value of a cycle of $\pi_k$'s which are all identical). 
In this regard we notice that the energy of the latter (\ref{noncompact})
is infinite in the stringy limit. It is understood that we are
using this limit only to identify the various distinct physical
configurations.

Let us first see how the IIA picture emerges in 
our framework, i.e. from properties of the gauge field
moduli space. 
To this end, we simply go back to the results of the previous section. 
Consider a sector of the $U(N)$  gauge theory
characterized by some definite twisted boundary
condition on the gauge fixed configuration (\ref{gaugefix}) for $A_\mu$, 
\be \label{resA}
A_\mu (\s+2\pi)= g A_\mu(\s) g^\da, 
\eq
where $g$ is a representative of a non--trivial conjugacy class in $S_N$. 
This induces the following twisted boundary condition on
$X^i_{mn}$,
\be \label{resX}
X^i (\s+2\pi)= g X^i (\s) g^\da.
\eq
We have already noticed that, since (\ref{gaugefix}) 
is independent of $\s$, (\ref{resA}) implies
\be 
\b_{g(i)} =\b_i.
\eq
The physical meaning of this sector is evident in the free string limit, 
(\ref{resX}) gives
\be 
X_{g(i)}(\s+2\pi) =X_i(\s).
\eq
These are long string configurations.
To be more specific, consider a particular sector of the gauge theory
specified by the conjugacy class 
\be 
[g] =(1)^{N_1} (2)^{N_2} \cdots (s)^{N_s},
\eq 
We are going to have
\be \label{k-wise}
\pi^i_{m_1} =\pi^i_{m_2} = \cdots \pi^i_{m_k}, \quad 
k=1, \cdots, s,\quad i= 1, \cdots, N_k
\eq
which gives rise to a set of
$N_k$ strings
of length $2\pi k$, $k=1,\cdots, s$. 
Intuitively, we can think of the process of `screwing' strings as follows.
A generic point in the gauge theory moduli space has 
$\pi_k$ all different. 
This corresponds to  $N$ strings with different electric flux circulating 
the strings.
Due to conservation of electric flux, two strings with
different flux cannot join. However, 
corresponding to the point (\ref{k-wise}) in the moduli space
for which a number of strings have the same flux living on them, they can
combine with each other and give rise to long strings.

What we have said so far about the ${\cal A}_F$ theory is not complete.
The `quantum moduli space' of $\mbox{SYM}_{1+1}$ is characterized 
by another quantum number, the common value of the $\pi_k$ in a given cycle,
which has already appeared in (\ref{k-wise}) above without interpretation.
We recall that the trace of the electric flux is usually interpreted as
a D0--brane charge, \cite{BFSS,flux,BSS}. 
To see more clearly the relation of theses D0--branes with the long strings, 
it is convenient to switch to a IIB picture. 
So we do a T--duality on ${\cal A}_D$ and land on ${\cal B}_D$. 
In a IIB picture we expect D--strings, elementary strings 
and bound states of them, with integral charges $(q_e,q_m)$.
We show now the SYM gauge moduli space has 
a natural interpretation in terms of these charges.
To see this, it is enough to look at a particular subsector of the
gauge theory, 
for example, a cycle of length $n$ with an `electric flux',
\be \label{Em}
\pi_1 =\pi_2 = \cdots = \pi_n =m , \quad m \geq 0.
\eq
This  corresponds to a long string of length $2 \pi n$ in the ${\cal A}_F$
picture, and according to T--duality, should correspond to a certain
D-string configuration. Indeed a long string of length $2 \pi n$ 
can be thought of as coming from $n$ fundamental
IIA strings. One can do a combined T--duality and  
S--duality operation. It is expected that 
this long string be mapped to 
$n$ D--strings in the ${\cal B}_D$ picture. 
One can also see this more intuitively by rescaling the long string 
back to standard world sheet dimension. Since the
total mass of the string is not changed, the tension will get 
increased by a factor of $n$. 
This can be identified with the tension of a D-string 
carrying RR-charge $n$ \cite{pol},
\be 
T \sim n/g_B.  
\eq

Next we identify (\ref{Em}) with a IIB string of type $(m,n)$.
One can compute the energy per unit length associated with ({\ref{Em}), 
\be \label{modifiedE}
\CE/L = \frac{\gy^2}{2n} m^2. 
\eq
This is, to lowest order,
exactly what one would expect $m$ 
fundamental IIB strings to contribute to the IIB $(m,n)$ string tension,
\be
 T_{(m,n)}= \sqrt {m^2 + \frac{n^2}{g_B^2}} 
 \sim \frac{n}{g_B}+ \frac{m^2}{2n}g_B. 
\eq

Therefore a long string of 
length $2 \pi n$, together with the electric flux
(\ref{Em}) can be mapped to a collection of $n$ D-strings together with
$m$ fundamental strings, i.e. a IIB $(m,n)$ string 
according to \cite{bdd}. 
 
Thus one can take the  two dimensional $U(N)$ SYM (\ref{SYM'}) 
as describing a collection of $(m,n)$ IIB strings, with 
$n \leq N$, stretched in the $9^{th}$ direction of 
radius $1/R_9$. 
The large $N$ limit will restore the full spectrum required by
the $SL(2,\Z)$  symmetry of IIB string theory. 
Stability of $(m,n)$ strings 
requires $m,n$ to be relatively prime, but this cannot be
seen solely from the properties of pure gauge sector and is related to
the detailed properties of the full supersymmetric 
system, \cite{bdd}.

Let us now switch back to ${\cal A}_F$. Since fundamental strings in
${\cal B}_D$ are mapped to D0--branes in ${\cal A}_F$, we see that an
$(m,n)$ sector of SYM corresponds to $m$ D0--branes attached to
a long string of length $2n\pi$.  
This is of course consistent. (\ref{Em}) gives an electric flux 
\be 
\frac{1}{\gy^2}tr E = \frac{1}{n} \sum_k \pi_k =m
\eq
of $m$ units and therefore represents  $m$ D0 branes.
The energy  $\CE = \frac{\gy^2}{2n} L m^2$ is also what one
would expect \cite{DVV1} from a configuration of $m$ D0 branes adhering to 
the long string.

Let us now comment on the inclusion of all the twisted sectors 
in the IIA string interpretation \cite{motl,BS,DVV1}. 
It is common in string orbifold
construction to include twisted sectors in the orbifold Hilbert space. 
Precisely which twisted sectors have to
be included depends on the physical problem. For ordinary string orbifolds,
modular invariance requires the inclusion of all the twisted sectors. 
Now, let us go back to our SYM. We just saw that the 
`long strings configurations'
can be identified with IIB $(m,n)$ strings. It is clear that from
the requirement of $SL(2, \Z)$ symmetry of type IIB theory, we should
include all of them, i.e. in ${\cal B}_D$ we should include all the $(m,n)$
subsectors. This also entails that the same sectors should be present on
the IIA side. In other words the $SL(2, \Z)$ symmetry of type IIB theory plays, 
in $\mbox{SYM}_{1+1}$ of M(atrix) theory compactifications, a role analogous 
to modular invariance in ordinary string orbifolds .

Finally, we comment briefly on ${\cal B}_F$. 
We pass from ${\cal B}_D$ to it via a 9--11 flip, 
which corresponds to an S--duality operation. 
Therefore D--strings are mapped to 
fundamental strings and vice versa and we have the
opposite assignment of the quantum numbers $n$ and $m$, 
with respect to above.
In other words, the original IIA long strings of ${\cal A}_F$
have become now the fundamental strings in ${\cal B}_F$ 
and are assigned the quantum number $n$. 
This is of course consistent, since the relation
between ${\cal B}_F$ and ${\cal A}_F$ is a T--duality operation, 
which leaves
unchanged the D-- or F--character of the string configurations.
In fact, (\ref{modifiedE}) in this case becomes $\frac{m^2}{2n g_B^2}$,
which agrees with what one would expect from the tension of a
$(n,m)$ string,
\be
 T_{(n,m)}=\sqrt{n^2 + \frac{m^2}{g_B^2} }\sim 
n+ \frac{m^2}{2n g_B^2}.
\eq

\section{Discussions}

Throughout this paper we have kept $N$, $R_9$ and $R_{11}$ finite. 
We have interpreted SYM as an approximate description of
9--dimensional theories, in particular (\ref{SYM'}) with (\ref{bd}) as an
effective description of D--strings. This theory 
was originally used by  Witten, \cite{bdd}, as an 
effective description of coincident IIB D-strings to study bound
state problems. Given the M/IIB duality (see for example \cite{schwarz}), 
one can think of it as a partial description of 
M theory compactified on a torus with radii $R_9, R_{11}$. 

We have seen above that we can recover the full IIB (and consequently also the
IIA) spectrum by taking $N\to \infty$. Therefore a discussion of this
limit is unavoidable. The best we can say is that in the framework of
M(atrix) theory, the issue of large $N$ limit seems far from 
being well established. On the one hand we can take the original attitude
of \cite{BFSS} and say that (\ref{qm}) in the $N\to \infty$ {\it and} 
$R_{11}\to\infty$ limit represents 11--dimensional M theory. 
On the other hand, we
can think of taking the $N\to \infty$ limit, while keeping $R_{11}$ finite
and ask ourselves whether this corresponds to any sensible theory (in 10D).
We do not have a decisive argument to choose this second attitude.
However, if we consider further compactification on a circle of 
radius $R_9$, as in the last section, given the nice interpretations 
of SYM presented there,
which agree very well with the notions we have about type IIA and type IIB
theories in 9--dimensions, we are oriented to assume that the $N\to \infty$
limits of the theories (\ref{SYM'}) represent enlarged IIA and IIB theories
in 9 dimensions. This seems somehow to imply that (\ref{qm}) represents in
the large $N$ limit M theory compactified on a circle.
 
Assuming this, (\ref{SYM'}), which was an effective description for 9D theory,
becomes in the large $N$ limit an exact description of M theory compactified 
on a torus. Thus, the two dimensional SYM provides a unified description
of both IIA and IIB. Let us comment about this possibility. 
According to \cite{schwarz}, the M theory membrane can wrap on a torus 
in different ways and give rise to the various IIB $(m,n)$ strings.
To recover 10 dimensional 
IIB, one should let $R_9, R_{11} \rightarrow 0$, with $g_B=R_{11}/R_9$
fixed. The choice (\ref{bd}) of $\gy$ in (\ref{SYM'}) is appropriate for this 
case. On the other hand we can take $R_{11}\to \infty$ and consider M theory
compactified on a circle, which must coincide with IIA. This corresponds
to the choice (\ref{af}) made in \cite{DVV1}, which is
suitable for this purpose, since $R_{11}$ does not appear in (\ref{SYM}).
 
\vskip.3cm
{\bf Acknowledgements.} We would like to acknowledge useful discussions we
had with Pei-Ming Ho and Cesare Reina. This research was partially 
supported by EC TMR Programme, grant FMRX-CT96-0012.

\vskip.3cm

{\bf Note added}: While this work was being typed, the paper \cite{HV} 
which also propose to treat the two dimensional SYM as describing IIB 
fundamental strings.

\section{Appendix}
In this appendix, we 
repeat the procedure of \cite{wati},
We start with the D0 brane action (\ref{qm}) and rescale $X^i$ so that
\be \label{r11}
S=\int dt L =\int dt \frac{1}{2 R_{11}^3} tr \left( 
\dot{X}^{i 2} - [X^i, X^j]^2 
+ \th^T \dot{\th} -\th^T \g_i [X^i,\th] \right).
\eq
Compactifing this  on a circle of radius $R_9$, we obtain
(without  explicitly writing the fermionic terms)
\be
S = \int dt L 
=\int dt \int_0^{2\pi/R_9} d\s \frac{R_9}{2 R_{11}^3}
tr \left( \dot{X}^{i 2} + \dot{A_1}^2 - (D_\s X^i)^2 
- [X^i, X^j]^2 \right). 
\eq
Rescaling the world sheet back to normal length $2 \pi$,
and also rescaling $A_1$ such that 
$D_\s \rightarrow R_9 D_\s$,
\be
S=\frac{1}{2}\int dt \int_0^{2\pi} d\s 
tr \left( \frac{ \dot{X}^{i 2} }{R_{11}^3}+
\frac{R_9^2}{R_{11}^3} \dot{A_1}^2 - 
\frac{R_9^2}{R_{11}^3} (D_\s X^i)^2  
-\frac{1}{R_{11}^3}  [X^i, X^j]^2. \right)
\eq
Now rescale the coordinates 
$X \rightarrow (R_{11}^3/R_9)^{1/2} X$ 
\bee \label{worldtime}
S &=&\frac{1}{2}\int dt \int_0^{2\pi } d\s 
tr \left( \frac{ \dot{X}^{i 2} }{R_9}
+\frac{R_9^2}{R_{11}^3} \dot{A_1}^2 
- R_9 (D_\s X^i)^2  
-\frac{R_{11}^3}{R_9^2} [X^i, X^j]^2 \right)\\
&:=& \int dt \int_0^{2\pi} d\s \CL. \n
\eqq

Using $\CL$, one can construct the corresponding 
Hamiltonian
\be
H= \frac{R_9}{2} \int_0^{2\pi} d\s   
tr \left( \Pi_i^2 + (D_\s X^i)^2  
+(\frac{R_{11}}{R_9})^3 (\Pi_A^2+ [X^i, X^j]^2)
\right)
\eq
where $\Pi_A$ is the conjugate momentum of $A_1$.
The gauge invariant form of the action can be obtained by
absorbing the overall factor of $R_9$ into the
definition of world sheet time in (\ref{worldtime}):
$t \to t/R_9$, 
\be
S = \frac{1}{2}\int dt \int_0^{2\pi } d\s 
tr \left(  (D_\mu X^i)^2  
+\frac{R_{9}^3}{R_{11}^3}F_{\mu \nu}^2 
-\frac{R_{11}^3}{R_9^3}[X^i, X^j]^2).
\right)
\eq
Different dependences of the Yang-Mills couplings, such as those of 
eqs.(\ref{ad}--\ref{bf}), can be obtained by
noticing that the explicit dependence of $R_{11}$ in (\ref{r11})
can be removed or, anyhow, modified by means of field redefinitions
together with a redefinition of time. 

We believe world--sheet time redefinitions are not irrelevant in defining
the appropriate large $N$ limit.


\begin{thebibliography}{}

\bibitem{motl}
L.~Motl, {\it Proposals on Nonperturbative Superstring Interactions},
{\tt hep-th/9701025}.

\bibitem{BS}
T.~Banks and N.~Seiberg,
{\it Strings from Matrices}, {\tt hep-th/9702187}.

\bibitem{DVV1}
R. Dijkgraaf,  E. Verlinde, H. Verlinde,
{\it Matrix String Theory},
{\tt hep-th/9703030}.


\bibitem{BFSS}
T. Banks, W. Fischler, S. H. Shenker and  L. Susskind,
{\it M Theory As A Matrix Model: A Conjecture},
{\tt hep-th/9610043}.

\bibitem{DMVV}
 R. Dijkgraaf, G, Moore, E. Verlinde, H. Verlinde,
{\it Elliptic Genera of Symmetric Products and Second Quantized Strings},
Commun. Math. Phys. to be published, {\tt hep-th/9608096}.
 
\bibitem{hos} J.E.Hetrick and Y.Hosotani 
{\it Yang--Mills Theory on a Circle}, Phys.Lett. {\bf B230} (1989), 88.

\bibitem{sem} E.Langmann and G.W.Semenoff, 
{\it Gauge Theory on a Cylinder}, Phys.Lett. {\bf B296} (1992), 117,
{\tt hep-th/9210011}.
{\it Gribov Ambiguity and Non--Trivial Vacuum 
Structure of Gauge Theories on a Cylinder}, 
Phys.Lett. {\bf B303} (1993), 303,{\tt hep-th/9212083}.

\bibitem{hetrick} J.E.Hetrick, 
{\it Canonical Quantization of Two Dimensional 
Gauge Fields}, Int. Jour. Mod. Phys. 
{\bf A9} (1994), 3153, {\tt hep-th/9305020}.

\bibitem{raj} S.G.Rajeev, {\it Yang--Mills Theory on a Cylinder}, Phys.Lett. 
{\bf B212} (1988), 203.


\bibitem{ulf}
U.H.~Danielsson, G.~Ferretti, and B.~Sundborg, {\it D-particle
Dynamics and Bound States}, {Int. J. Mod. Phys.}  {\bf A11}
(1996) 5463--5478, {\tt hep-th/9603081};\\
 D.~Kabat and P.~Pouliot, {\it A Comment on
Zerobrane Quantum Mechanics}, Phys. Rev. Lett. {\bf 77} (1996)
1004--1007, {\tt hep-th/9603127};\\
M.~R. Douglas, D.~Kabat, P.~Pouliot, and S.~H. Shenker, 
{\it D-branes and Short Distances in String Theory}, 
Nucl. Phys. {\bf B485} (1997) 85--127,
{\tt hep-th/9608024}.

\bibitem{wati}
W.~Taylor, {\it D-brane Field Theory on Compact Spaces},
 {\tt hep-th/9611042}.

\bibitem{HW}
P.M.~Ho and  Y..S.~Wu, {\it IIB/M Duality and Longitudinal Membranes},
{\tt hep-th/9703016.}

\bibitem{ST}
J.A.~Harvey, G.~Moore, and A.~Strominger, {\it Reducing $S$-Duality to
$T$-Duality}, Phys.Rev. {\bf D52} (1995) 7161,
{\tt hep-th/9501022}.\\
 M. Bershadsky, A. Johansen, V. Sadov, C. Vafa,
{\it Topological Reduction of 4D SYM to 2D $\sigma$--Model},
Nucl. Phys. {\bf B448} (1995) 166, {\tt hep-th/9501096}.

\bibitem{flux}
O. J. Ganor, S. Ramgoolam and W. Taylor,
{\it Branes, Fluxes and Duality in M(atrix)-Theory},
{\tt hep-th/9611202}.

\bibitem{BSS}
T.Banks, N.Seiberg and S.Shenker, {\it Branes from Matrices}, 
{\tt hep-th/9612157}.

\bibitem{pol} J. Polchinski, {\it  Dirichlet-Branes and Ramond-Ramond
Charges}, {\tt hep-th/951001.}

\bibitem{bdd}
E.~Witten, {\it Bound {S}tates of {S}trings and $p$-{B}ranes}, 
Nucl. Phys. {\bf B460} (1996) 335,
{\tt hep-th/9510135}.


\bibitem{schwarz}
J.~H. Schwarz, {\it Lectures on Superstring and M Theory Dualities},
{\tt hep-th/9607201}.
 
  
\bibitem{HV}
H.Verlinde,
{\it A Matrix String Interpretation of the Large N Loop Equation},
{\tt hep-th/9705029}.

\end{thebibliography}
\end{document}